\documentclass{article}
\pdfoutput=1
\usepackage{arxiv}

\usepackage[utf8]{inputenc} % allow utf-8 input
\usepackage[T1]{fontenc}    % use 8-bit T1 fonts
\usepackage{hyperref}       % hyperlinks
\usepackage{url}            % simple URL typesetting
\usepackage{booktabs}       % professional-quality tables
\usepackage{amsfonts}       % blackboard math symbols
\usepackage{nicefrac}       % compact symbols for 1/2, etc.
\usepackage{microtype}      % microtypography
\usepackage{lipsum}

\usepackage{amsmath}
\usepackage{graphicx,psfrag,epsf}
\usepackage{enumerate}
\usepackage{natbib}

\usepackage{adjustbox}
\usepackage{multirow}
\usepackage{xcolor}
\usepackage{caption}
\usepackage{subcaption}
\usepackage{chngcntr}
\usepackage[flushleft]{threeparttable}

\renewcommand{\theequation}{\arabic{equation}}

\title{Causal Inference on Win Ratio for Observational Data with Dependent Subjects}

\author{
Di Zhang\\
Real World Evidence Statistics\\
Teva Pharmaceutical\\
West Chester, Pennsylvania, USA\\
\And
Stephen R. Wisniewski\\
Department of Epidemiology\\
Graduate School of Public Health\\
University of Pittsburgh, Pittsburgh, USA\\
\And
Jong-Hyeon Jeong \thanks{Corresponding Author: jjeong@pitt.edu}\\
Department of Biostatistics\\
Graduate School of Public Health\\
University of Pittsburgh, Pittsburgh, USA\\
}

\begin{document}
\maketitle

\begin{abstract}
Composite endpoints are commonly used with an anticipation that clinically relevant endpoints as a whole would yield meaningful treatment benefits. The win ratio is a rank-based statistic to summarize composite endpoints, allowing prioritizing the important components of the composite endpoints. Recent development in statistical inference for the win ratio statistic has been focusing on independent subjects without any potential confounding. When analyzing composite endpoints using observational data, one of the important challenges is confounding at baseline. Additionally, hierarchical observational data structures are commonly seen in practice, especially in multi-center studies with patients nesting within hospitals. Such hierarchical structure can introduce potential dependency or cluster effects among observations in the analysis. To address these two issues when using the win ratio statistic, we propose a weighted stratified causal win ratio estimator with calibrated weights. The calibrated weights create balanced patient-level covariates and cluster effect distributions between comparison groups. We conducted extensive simulation studies and showed promising performance of the proposed estimator in terms of bias, variance estimation, type I error and power analysis, regardless of the allocation of treatment assignments at baseline and intra-cluster correlations within clusters. Lastly, the proposed estimator was applied to an observational study among children with traumatic brain injury.
\end{abstract}

\section{Introduction} \label{intro}
Composite endpoints are commonly used in clinical trials and observational studies. One of the main advantages of using the composite endpoints is a higher study power compared to a study design based on a single endpoint \cite{huque2011addressing}. When the composite endpoints imply different clinical importance, however, the analysis should take into account the magnitude of importance of each endpoint to reach a more sensible conclusion. One approach to account for the importance of composite endpoints is the win ratio statistic proposed by \cite{pocock2011win}. Without loss of generality, we consider composite endpoints of a non-terminal (e.g. hospitalization) and a terminal (e.g. death) event. To construct the win ratio statistic, time to the terminal events are first being compared between patients from treatment and control groups. If the “win” or “loss” decision cannot be made due to censoring, then the non-terminal events are compared subsequently to reach a decision. The verdict is a tie when at least one of the paired times is censored and one censoring time occurs prior to the other censoring time or event time. Eventually, the win ratio for the treatment group is the ratio of the number of wins to the number of losses in the treatment group. 
\\ \\
Numerous studies have investigated the statistical properties of the win ratio statistic in clinical trial settings. \cite{luo2015alternative}, \cite{dong2016generalized} and \cite{oakes2016win} established the statistical framework for the win ratio. \cite{bebu2015large} formulated the win ratio through a bivariate $U$-statistics for independent data based on large sample inference. \cite{luo2017weighted} proposed a weighted win ratio statistic to improve efficiency. Several studies have also extended the use of the win ratio to non-time-to-event outcomes \citep{wang2016win}. \cite{dong2019win}, and \cite{finkelstein2019graphing} discussed the interpretation of the win ratio. \cite{dong2018stratified} and \cite{mao2019alternative} further generalized the use of the win ratio to multivariate and stratified settings. 
\\ \\
The application of the win ratio statistic in observational studies raises additional challenges. The Approaches and Decisions for Acute Pediatric (ADAPT) Traumatic Brain Injury (TBI) trial is a multi-center observational study to investigate the impacts of various medical interventions on children under 18 years old with TBI. During the first 7 days of hospitalization after brain injuries, one of the study interests is to investigate the impact of the Cerebrospinal Fluid (CSF) drainage treatment compared to no CSF treatment on the composite endpoints of time to first most severe neurological complication and time to death. Patients and physicians value death as a more important event than complications, and the investigators want to reflect this information in the analysis. Total 26 potential baseline confounders were identified in the study, and the data were collected from 37 sites. The medical resources and physician training may be similar within the same site; however, they could vary substantially across sites. These challenges should be accounted for in the analysis to allow legitimate  inference. 
\\ \\
To adopt the win ratio statistic in the ADAPT observational study, we face two main challenges: potential confounders and cluster effects due to sites. The existing literature related to the inference of win ratio statistic has been mainly focusing on independent subjects without potential confounding. Mao proposed an efficient inverse probability weighted (IPW) estimator to make causal inference on the U-statistics \citep{mao2017causal}. In epidemiological literature, the approach of IPW type estimators in casual inference is also called propensity score (PS) analysis with weighting \citep{austin2015moving}. The propensity scores can be used in various ways to balance the baseline confounding covariates. Since the win ratio can be expressed as the rank-based U-statistic, in this paper we adopt Mao's efficient estimator for causal inference on the win ratio statistic to account for the baseline confounding. In terms of accounting for the cluster effects in observational studies, existing literature has been mainly focusing on the PS-based approaches. \cite{li2013propensity} and \cite{thoemmes2011use} have shown that one should account for the cluster effects both at the PS estimation stage and the outcome analysis stage. \cite{arpino2016propensity} proposed a PS matching algorithm with the PS estimation from fixed effects and random effects models. However, there are limitations based on both fixed effects and random effects models to estimate the PS \citep{li2013propensity} due to parametric assumptions. In addition, the cluster effects estimated from the random effects model can be distorted due to the estimation shrinkage towards the grand mean in the generalized mixed model. \cite{chan2016globally} introduced a general class of the calibration estimator for the weights to balance the baseline covariates for the independent subject case. This calibration estimation is attractive due to the non-parametric nature and its robustness against model misspecification. The calibration weighting technique is commonly used in survey sampling analysis \citep{chen2002using, kim2010calibration}, and recently, \cite{yang2018propensity} adapted this technique to estimate the weights for clustered continuous and binary outcomes.
\\ \\
We propose a weighted stratified win ratio estimator to analyze the composite endpoints in the presence of baseline confounding and cluster effects. We adopt the causal U-statistics by \cite{mao2017causal} to account for baseline confounding and the calibrated weights by \cite{yang2018propensity} to account for cluster effects in the treatment selection process. The proposed estimator was implemented in an R function available on GitHub. This article contributes to the methodological gap in the literature and in practice, regarding analyzing hierarchical observational data with subject-level confounding for composite endpoints, prioritizing important components of the composite endpoints. The rest of the article is organized as follows. Section \ref{wr_independent_no_confounding} reviews the win ratio statistic for independent subjects without confounding. Section \ref{wr_independent_confounding} presents the inference on the causal win ratio for independent subjects. Section \ref{weighted_estimator} introduces the proposed weighted stratified estimator for dependent subjects with confounding. Section \ref{weight_estimation} presents the estimation of the calibrated weights. Through simulation studies in Section \ref{calibrationWR_simulation}, we show that the proposed estimator performs well compared to the estimators with the weights estimated from the fixed effects or random effect models. An application to the TBI data is presented in Section \ref{calibrationWR_example}. Conclusions and limitations are discussed in Section \ref{calibrationWR_conclusion}.

%%%%%%%%%%%%%%%%%%%%%%%%%%%%%%%%%%%%%%%%%%%%%%%%%%%%%%%%%%%%%%%%%%%%%%%%%%%%%%%%%%%%%%%%%%%%%%%%%%%%%%%%%%%%%%%%%%%%%%%%%%%%%%%%%%%%%%%%%%%%%%%%%%
%%%%%%%%%%%%%%%%%%%%%%%%%%%%%%%%%%%%%%%%%%%%%%%%%%%%%%%%%%%%%%%%%%%%%%%%%%%%%%%%%%%%%%%%%%%%%%%%%%%%%%%%%%%%%%%%%%%%%%%%%%%%%%%%%%%%%%%%%%%%%%%%%%
%%%%%%%%%%%%%%%%%%%%%%%%%%%%%%%%%%%%%%%%%%%%%%%%%%%%%%%%%%%%%%%%%%%%%%%%%%%%%%%%%%%%%%%%%%%%%%%%%%%%%%%%%%%%%%%%%%%%%%%%%%%%%%%%%%%%%%%%%%%%%%%%%%
%%%%%%%%%%%%%%%%%%%%%%%%%%%%%%%%%%%%%%%%%%%%%%%%%%%%%%%%%%%%%%%%%%%%%%%%%%%%%%%%%%%%%%%%%%%%%%%%%%%%%%%%%%%%%%%%%%%%%%%%%%%%%%%%%%%%%%%%%%%%%%%%%%
\section{Review Win Ratio Statistic}
\label{wr_independent_no_confounding}
We first review the win ratio statistic for data with independent subjects without confounding \cite{pocock2011win}. Consider composite endpoints including a terminal event (e.g. death) and a non-terminal event (e.g. hospitalization). To decide if a treatment patient wins or loses compared to a control patient, times to the terminal events are first compared. If the control patient has an earlier terminal event, then the treatment patient wins, vice versa. If a decision cannot be reached using the terminal events due to censoring, then times to non-terminal events are compared between these two patients following the same logic. Similar procedures apply for all pairwise comparisons between patients from treatment and control groups. All possible scenarios of between-patient comparisons can be represented using the indicator functions as shown by \cite{luo2015alternative}. Let $N_w$ and $N_l$ be the numbers of wins and losses in the treatment group, respectively. The win ratio statistic for independent data is defined as $WR=E(\frac{N_w}{N_l})$, which can be approximated by $WR\approx \frac{E(N_w)}{E(N_l)}=\frac{\tau_1}{\tau_2}$ based on the first order Taylor series expansion.

%%%%%%%%%%%%%%%%%%%%%%%%%%%%%%%%%%%%%%%%%%%%%%%%%%%%%%%%%%%%%%%%%%%%%%%%%%%%%%%%%%%%%%%%%%%%%%%%%%%%%%%%%%%%%%%%%%%%%%%%%%%%%%%%%%%%%%%%%%%%%%%%%%
%%%%%%%%%%%%%%%%%%%%%%%%%%%%%%%%%%%%%%%%%%%%%%%%%%%%%%%%%%%%%%%%%%%%%%%%%%%%%%%%%%%%%%%%%%%%%%%%%%%%%%%%%%%%%%%%%%%%%%%%%%%%%%%%%%%%%%%%%%%%%%%%%%
%%%%%%%%%%%%%%%%%%%%%%%%%%%%%%%%%%%%%%%%%%%%%%%%%%%%%%%%%%%%%%%%%%%%%%%%%%%%%%%%%%%%%%%%%%%%%%%%%%%%%%%%%%%%%%%%%%%%%%%%%%%%%%%%%%%%%%%%%%%%%%%%%%
%%%%%%%%%%%%%%%%%%%%%%%%%%%%%%%%%%%%%%%%%%%%%%%%%%%%%%%%%%%%%%%%%%%%%%%%%%%%%%%%%%%%%%%%%%%%%%%%%%%%%%%%%%%%%%%%%%%%%%%%%%%%%%%%%%%%%%%%%%%%%%%%%%
\section{Win Ratio for Independent Subjects with Confounding}
\label{wr_independent_confounding}
Based on the formulation of independent win ratio by \cite{luo2015alternative} and \cite{bebu2015large}, we adopt the efficient estimator for the causal U-statistics by \cite{mao2017causal} to account for patient-level confounding. The win ratio statistic is essentially a ratio of two ordinal outcomes and they can be represented by the U-statistics. \cite{mao2017causal} established the causal inference for the U-statistics using inverse probability weights (IPWs) based on the semi-parametric theory. We define $n$ to be the total number of independent subjects. Let $Y_i$ and $Y_j$ be the observed composite endpoints for subject $i$ and $j$, respectively, where $i, j=1,...,n$. We define two kernels as, $i\ne j$,
\begin{equation}\label{kernel_causalWR1}
\phi_1(Y_i,Y_j)=\textbf{1}\{Y_i>Y_j\}=
\left\{
\begin{array}{@{}ll@{}}
1 & \text{if}\ Y_i \text{ wins}, \\
0 & \text{otherwise},
\end{array}\right.
\end{equation}
and
\begin{equation}\label{kernel_causalWR2}
\phi_2(Y_i,Y_j)=\textbf{1}\{Y_i<Y_j\}=
\left\{
\begin{array}{@{}ll@{}}
1 & \text{if}\ Y_j \text{ wins}, \\
0 & \text{otherwise}.
\end{array}\right.
\end{equation}
The within-pair comparisons to determine whether $Y_i$ or $Y_j$ wins follow the description in Section \ref{wr_independent_no_confounding}. An efficient IPW estimator \citep{mao2017causal} can be constructed as below, using the observed data $Y_i$ and $Y_j$ in the kernel function (\ref{kernel_causalWR1})
\begin{equation}
\label{ipw}
\begin{split}
\hat{\tau_1}=&{n \choose 2}^{-1} \sum_{i=1}^{n-1}\sum_{j>i}^{n}\{\frac{1}{2}\{\frac{Z_i(1-Z_j)}{\hat{\pi} (X_i; \hat{\alpha})(1-\hat{\pi}(X_j, \hat{\alpha}))}\phi_1(Y_i, Y_j)\\
&+\frac{Z_j(1-Z_i)}{\hat{\pi}(X_j; \hat{\alpha})(1-\hat{\pi}(X_i; \hat{\alpha}))}\phi_1(Y_j, Y_i)\}\},
\end{split}
\end{equation}
where $Z_i$ is the observed treatment status for subject $i$ and $\pi(X_i; \alpha)$ is the probability of receiving the treatment of interest given a vector of confounders $X_i$ for subject $i$. The only quantity that needs to be estimated in $\hat{\tau_1}$ is $\pi(\cdot)$ and it's usually done through some parametric modeling, such as a logistic regression with a model parameter vector $\alpha$. This average effect for the ordinal outcomes is defined as average superiority effect (ASE) \citep{mao2017causal}. Similar formulation can be applied for $\tau_2$. \cite{mao2017causal} showed that the IPW estimator for the average treatment effect in (\ref{ipw}) is consistent and asymptotically normally distributed under some suitable regularity conditions. Therefore, as $n \rightarrow \infty$, by the delta method the estimator of the causal win ratio, $\hat{\tau_1}/\hat{\tau_2}$, would also follow an asymptotic normal distribution as does $\log[\hat{\tau_1}/\hat{\tau_2}]$.

%%%%%%%%%%%%%%%%%%%%%%%%%%%%%%%%%%%%%%%%%%%%%%%%%%%%%%%%%%%%%%%%%%%%%%%%%%%%%%%%%%%%%%%%%%%%%%%%%%%%%%%%%%%%%%%%%%%%%%%%%%%%%%%%%%%%%%%%%%%%%%%%%%
%%%%%%%%%%%%%%%%%%%%%%%%%%%%%%%%%%%%%%%%%%%%%%%%%%%%%%%%%%%%%%%%%%%%%%%%%%%%%%%%%%%%%%%%%%%%%%%%%%%%%%%%%%%%%%%%%%%%%%%%%%%%%%%%%%%%%%%%%%%%%%%%%%
%%%%%%%%%%%%%%%%%%%%%%%%%%%%%%%%%%%%%%%%%%%%%%%%%%%%%%%%%%%%%%%%%%%%%%%%%%%%%%%%%%%%%%%%%%%%%%%%%%%%%%%%%%%%%%%%%%%%%%%%%%%%%%%%%%%%%%%%%%%%%%%%%%
%%%%%%%%%%%%%%%%%%%%%%%%%%%%%%%%%%%%%%%%%%%%%%%%%%%%%%%%%%%%%%%%%%%%%%%%%%%%%%%%%%%%%%%%%%%%%%%%%%%%%%%%%%%%%%%%%%%%%%%%%%%%%%%%%%%%%%%%%%%%%%%%%%
\section{Win Ratio for Dependent Subjects with Confounding}
\label{weighted_estimator}
We propose to use a weighted stratified causal win ratio estimator to account for potential baseline confounding and cluster effects. The stratification unit is a cluster. We re-define some notations in this section. Let $\tau_{1i}$ be the average effect of treatment in cluster $i$, and $\tau_{2i}$ be the average effect of control in cluster $i$ ($i=1, ..., m$). Define the cluster-specific win ratio for cluster $i$ be $\mu_i = \frac{\tau_{1i}}{\tau_{2i}}$. To estimate $\tau_{1i}$ accounting for the patient-level confounding, we adopt the efficient estimator for the causal U-statistic by \cite{mao2017causal}, shown in Section \ref{wr_independent_confounding}. The estimator $\hat{\tau}_{1i}$ for cluster $i$ can be defined as 
\begin{equation}
\label{cluster_tao}
\hat{\tau}_{1i}={n_i \choose 2}^{-1} \sum_{j=1}^{n_i-1}\sum_{j'>j}^{n_i}\{\frac{1}{2}\{Z_{ij}(1-Z_{ij'})\hat{w}_{ij}\hat{w}_{ij'}\phi_1(Y_{ij}, Y_{ij'})+Z_{ij'}(1-Z_{ij})\hat{w}_{ij}\hat{w}_{ij'}\phi_1(Y_{ij'}, Y_{ij})\}\}, 
\end{equation}
where $n_i$ is the total number of independent subjects in cluster $i$, $Z_{ij}$ and $Z_{ij'}$ are treatment status for subject $j$ and $j'$ ($j, j'=1, ..., n_i$) in cluster $i$, respectively, and $\hat{w}_{ij}$ and $\hat{w}_{ij'}$ are the estimated weights for subject $j$ and $j'$. The weight is a function of patient-level confounders and cluster effects. Under the independent subjects setting, the weight corresponds to the inverse of probability of receiving the observed treatment, given patient-level confounders. As shown in Equation (\ref{ipw}), when subject $i$ receives treatment and subject $j$ receives control, $Z_i=1$ and $Z_j=0$. The weights correspond to $\frac{1}{\hat{\pi} (X_i; \hat{\alpha})(1-\hat{\pi}(X_j, \hat{\alpha}))}$. Under the dependent subjects setting, we incorporate the calibrated weights as proposed by \cite{yang2018propensity} to further account for cluster effects. More details of the calibrated weights are presented in Section \ref{weight_estimation}. Let $Y_{ij}$ and $Y_{ij'}$ be the composite endpoints for subjects $j$ and $j'$ in cluster $i$, respectively. The two kernels, where $j\ne j'$, become
\begin{equation}\label{kernel_causalWR3}
\phi_1(Y_{ij},Y_{ij'})=\textbf{1}\{Y_{ij}>Y_{ij'}\}=
\left\{
\begin{array}{@{}ll@{}}
1 & \text{if}\ Y_{ij} \text{ wins}, \\
0 & \text{otherwise},
\end{array}\right.
\end{equation}
and
\begin{equation}\label{kernel_causalWR4}
\phi_2(Y_{ij},Y_{ij'})=\textbf{1}\{Y_{ij}<Y_{ij'}\}=
\left\{
\begin{array}{@{}ll@{}}
1 & \text{if}\ Y_{ij'} \text{ wins}, \\
0 & \text{otherwise}.
\end{array}\right.
\end{equation}
Similar estimator can be derived for $\tau_{2i}$ with the kernel $\phi_2(Y_{ij},Y_{ij'})$. 
\\ \\
Define $\hat{w}_{i\cdot}=\sum_{j=1}^{n_i-1}\sum_{j'>j}^{n_i}\{ \frac{1}{2} Z_{ij}(1-Z_{ij'})\hat{w}_{ij}\hat{w}_{ij'}\}$ as sum of all possible pairwise combinations of estimated weights multiplied for cluster $i$. Let $\hat{w}_{\cdot \cdot}=\sum_{i=1}^{m} \hat{w}_{i\cdot}$ be the overall estimated weights across all clusters. Therefore, the estimated average treatment effect of the causal win ratio for clustered data is
\begin{equation}
\label{equ7}
\hat{\mu} = \frac{\sum_{i=1}^{m} \hat{w}_{i\cdot}\hat{\mu}_i}{\hat{w}_{\cdot \cdot}}.
\end{equation}
Heuristically, assuming that $\hat{w}_{i \cdot}$ ($i=1,2,...,m$) and $\hat{w}_{\cdot \cdot}$ converge in probability to their limiting values $\omega_i$ ($i=1,2,...,n_i$) and $\omega$, as $m, n_i \rightarrow \infty$, $\hat{\mu}$ would follow an asymptotic normal distribution by the Slutsky theorem as $\hat{\mu}_i$ does. In practice, the variance of this estimator can be estimated using empirical bootstrap variance estimation by resampling clusters with replacement. We first draw samples of $m$ clusters independently with replacement. Then estimate the causal win ratio as shown in Equation (\ref{equ7}) in this bootstrap sample. Repeat this resampling and estimation procedure for $B$ times. The bootstrap variance is the sample variance of the causal win ratio estimates from the bootstrap samples. We show the empirical distribution of the proposed estimator follows a normal distribution in the Appendix, based on our simulation set-up in Section \ref{set-up}. As the number of clusters and the cluster sizes increase, the density of the distribution concentrates more around the mean. 
\\ \\
Under the potential outcomes framework \citep{rubin1974estimating}, we assume that the consistency assumption holds true, which defines that the observed outcome corresponds to the potential outcome under the specific treatment. Furthermore, we make the stable unit and treatment version assumption (SUTVA), which implies that the potential outcomes for each unit are not affected by the treatments assigned to other units. Conditional on the observed patient-level confounders and cluster effects, the ignoreability assumption holds, where the potential outcomes and treatment variable are independent. Yang defines this assumption as latent ignorability assumption for cluster data \citep{yang2018propensity}. Lastly, we make positivity assumption that every unit has a positive probability to receive the treatment of interest. This is reflected in $0<\pi(X_{ij}; \alpha)<1$ for the initial weight estimation in Section \ref{weight_estimation}.

%%%%%%%%%%%%%%%%%%%%%%%%%%%%%%%%%%%%%%%%%%%%%%%%%%%%%%%%%%%%%%%%%%%%%%%%%%%%%%%%%%%%%%%%%%%%%%%%%%%%%%%%%%%%%%%%%%%%%%%%%%%%%%%%%%%%%%%%%%%%%%%%%%
%%%%%%%%%%%%%%%%%%%%%%%%%%%%%%%%%%%%%%%%%%%%%%%%%%%%%%%%%%%%%%%%%%%%%%%%%%%%%%%%%%%%%%%%%%%%%%%%%%%%%%%%%%%%%%%%%%%%%%%%%%%%%%%%%%%%%%%%%%%%%%%%%%
%%%%%%%%%%%%%%%%%%%%%%%%%%%%%%%%%%%%%%%%%%%%%%%%%%%%%%%%%%%%%%%%%%%%%%%%%%%%%%%%%%%%%%%%%%%%%%%%%%%%%%%%%%%%%%%%%%%%%%%%%%%%%%%%%%%%%%%%%%%%%%%%%%
%%%%%%%%%%%%%%%%%%%%%%%%%%%%%%%%%%%%%%%%%%%%%%%%%%%%%%%%%%%%%%%%%%%%%%%%%%%%%%%%%%%%%%%%%%%%%%%%%%%%%%%%%%%%%%%%%%%%%%%%%%%%%%%%%%%%%%%%%%%%%%%%%%

\section{Calibrated Weight Estimation}\label{weight_estimation}
We use calibrated weights, proposed by \cite{yang2018propensity}, to account for potential subject-level confounders and cluster effects imposed on treatment selections in observational studies. Following the notations in \cite{yang2018propensity}, there are two steps in the estimation procedure:
\begin{enumerate}
	\item Estimate the initial weights $d_{ij}$ using some parametric working model, such as the logistic regression. Define $d_{ij}=\frac{Z_{ij}}{\pi(X_{ij}; \alpha)}+\frac{1-Z_{ij}}{1-\pi(X_{ij}; \alpha)}$ as the initial weights, where $\pi(X_{ij}; \alpha)$ is the probability of receiving treatment of interest conditional on confounders for subject $j$ in cluster $i$, and $\alpha$ is a vector of coefficient parameters for the parametric working model. 
	\item Estimate the calibration weights $w_{ij}$ using some distance function $D(w_{ij}, d_{ij})$ with constraints. Define $w_{ij}$ as the final weight.
\end{enumerate}
The constrains in step 2 of the procedure are used to ensure the balance of covariates and cluster effects between comparison groups so that valid causal inference can be made for observational clustered data. We refer to \cite{yang2018propensity} for more detailed discussion on the constrains. Different distance functions can lead to different estimation methods. Following \cite{yang2018propensity}, we adopted the Kullback-Leibler (KL) distance function $D(w_{ij}, d_{ij})=w_{ij}\ln \frac{w_{ij}}{d_{ij}}$. This distance function measures distance between probability distributions. The KL distance function provides exponential tilting estimation \citep{schennach2007point}, which is asymptotically equivalent to the regression estimator, but avoids extreme weights \citep{kim2010calibration, schennach2007point}. Specifically in this problem, the KL distance function can incorporate the constraints as a closed form, which is more computationally efficient. 
\\ \\
The calibrated weights can be estimated by solving the distance function with constrains through iterative optimization. A detailed derivation of the calibrated weights following \cite{yang2018propensity} was provided in the Appendix. We implemented the proposed weighted stratified estimator with calibrated weights in a R function \emph{WR.causal} of package \verb|cWR| on GitHub at \url{https://github.com/dee1008/cWR}.

%%%%%%%%%%%%%%%%%%%%%%%%%%%%%%%%%%%%%%%%%%%%%%%%%%%%%%%%%%%%%%%%%%%%%%%%%%%%%%%%%%%%%%%%%%%%%%%%%%%%%%%%%%%%%%%%%%%%%%%%%%%%%%%%%%%%%%%%%%%%%%%%%%
%%%%%%%%%%%%%%%%%%%%%%%%%%%%%%%%%%%%%%%%%%%%%%%%%%%%%%%%%%%%%%%%%%%%%%%%%%%%%%%%%%%%%%%%%%%%%%%%%%%%%%%%%%%%%%%%%%%%%%%%%%%%%%%%%%%%%%%%%%%%%%%%%%
%%%%%%%%%%%%%%%%%%%%%%%%%%%%%%%%%%%%%%%%%%%%%%%%%%%%%%%%%%%%%%%%%%%%%%%%%%%%%%%%%%%%%%%%%%%%%%%%%%%%%%%%%%%%%%%%%%%%%%%%%%%%%%%%%%%%%%%%%%%%%%%%%%
%%%%%%%%%%%%%%%%%%%%%%%%%%%%%%%%%%%%%%%%%%%%%%%%%%%%%%%%%%%%%%%%%%%%%%%%%%%%%%%%%%%%%%%%%%%%%%%%%%%%%%%%%%%%%%%%%%%%%%%%%%%%%%%%%%%%%%%%%%%%%%%%%%

\section{Simulation Studies}
\label{calibrationWR_simulation}

\subsection{Set-up}
\label{set-up}
Several simulation studies were conducted to assess the behavior of the proposed weighted stratified causal win ratio estimator in clustered settings. The following estimators were compared, (1) the independent causal win ratio estimator without stratification and with IPW estimated from standard logistic regression (i.e. unadjusted estimator); (2) the weighted stratified estimator with IPW estimated from standard logistic regression (i.e. logistic estimator); (3) the weighted stratified estimator with IPW estimated from logistic regression with clusters as fixed effects (i.e. fixed estimator); (4) the weighted stratified estimator with IPW estimated from logistic regression with clusters as random intercepts (i.e. random estimator); (5) the weighted stratified estimator with calibrated weights (i.e. calibration estimator). In the calibrated weight estimation, we used a standard logistic regression as the working model to estimate the initial weights $d_{ij}$. Since the statistical framework we assumed requires both treatment and control groups within the same cluster, we excluded clusters that only contained one group in the simulation. Two covariates were considered, $X_1$ was a standard normal variable, and $X_2$ was normally distributed with mean 1 and variance 4. The same set of covariates were used to generate the outcomes and the treatments. Let $\gamma_1$ be the cluster effect on the outcomes and $\gamma_2$ be the cluster effect on the treatment selections. We assume $\gamma_1$ follows a gamma distribution that induces the intra-cluster correlation (ICC) of 0.2 or 0.067. We assume that $\gamma_2$ follows a normal distribution with the mean of 0 and the variance of 4 or a gamma distribution with the shape parameter of 2 and the rate parameter of 10. We assume $\gamma_1$ and $\gamma_2$ are correlated through a normal copula with a correlation parameter of 0.4. The cluster effects were generated using the R package \verb|Copula| \citep{kojadinovic2010modeling}. 
\\ \\
Assuming a proportional hazards model, let $\lambda_{H_{z_{ij}}}=\gamma_{1i}\lambda_H\exp(-\eta_{H}z_{ij}+x_{ij}\beta_1)$ be the hazard function for time to non-terminal event given covariates for patient $j$ in cluster $i$, where $\lambda_H$ was the baseline hazard function of the distribution of time to non-terminal events, $\eta_H$ was the non-terminal event rate per time unit, and $\beta_1$ was the coefficients of the covariate vector $x_{ij}$ with dimension 2. The treatment status $z_{ij}$ can be 0 or 1, which corresponds to treatment or control group, respectively. Similarly, let $\lambda_{D_{z_{ij}}}=\gamma_{1i}\lambda_D\exp(-\eta_{D}z_{ij}+x_{ij}\beta_2)$ be the hazard function of the distribution of time to terminal events. The bivariate exponential distribution with Gumbel-Hougaard copula was used to generate the bivariate distribution of time to non-terminal events $T_{H_{z_{ij}}}$ and time to terminal events $T_{D_{z_{ij}}}$ with the joint survival function
\begin{equation}
P(T_{H_{z_{ij}}}>y_1, T_{D_{z_{ij}}}>y_2|z_{ij}, x_{ij}, \gamma_{1i})=\exp\{-[(\lambda_{H_{z_{ij}}}y_1)^\varphi+(\lambda_{D_{z_{ij}}}y_2)^\varphi]^\frac{1}{\varphi}\},
\end{equation}
where $\varphi\ge 1$ is an association parameter between $T_{H_{z_{ij}}}$ and $T_{D_{z_{ij}}}$. Specifically to structure semi-competing risks data with covariates, the R package \verb|Gumbel| \citep{caillat2009package} was modified, such that the bivariate uniform random variables were first generated for the marginal survival functions in the copula based on the stable distribution \citep{marshall1988families}. Through the probability integral transformation, the true bivariate event times were obtained, adjusted for event types, covariates, and treatment effect. We assumed that the censoring time $T_{C_{z_{ij}}}$ was independent from $T_{H_{z_{ij}}}$, $T_{D_{z_{ij}}}$ and covariates, and followed an exponential distribution with the hazard function of $\lambda_{C_{z_{ij}}}=\lambda_C\exp(-\eta_{C}z_{ij})$. Treatment indicators were generated from a logistic regression, where $p(z_{ij}=1|x_{ij}, \gamma_{2i})=\frac{1}{1+\exp(-x_{ij}\alpha+\gamma_{2i})}$. 
\\ \\
The total sample size was 1000. We varied the number of clusters to be 20 or 50, and cluster sizes to be 50 or 20. Different percentage of treatment assignments were considered: 50\% (equal) and 30\% (unequal). Throughout the simulation study, the true parameter values were fixed as follows: $\lambda_H=0.1$, $\lambda_D=0.08$, $\lambda_C=0.09$, $\eta_C=0.1$, $\varphi=2$, $\beta_1=(0.1, 0.3)^T$, $\beta_2=(0.2, 0.4)^T$, $\alpha=(-0.2, 0.5, 0.5)^T$ for equal percentage of treatment assignment, and $\alpha=(-1.8, 0.5, 0.5)^T$ for unequal percentage of treatment assignment, assuming a normal distribution for the cluster effects on treatment selections. When the cluster effects on treatment selections follow the gamma distribution, $\alpha=(-0.6, 0.5, 0.5)^T$ for equal percentage of treatment assignment. Other parameters stay the same. To assess type I error probabilities, the data were simulated under null where $\log(\mu)=0$. For the power analysis, under the null hypothesis of $\log(\mu)=0$, the data were simulated from three different alternative values of $\log(\mu)=0.098$, 0.223, and 0.328, respectively. Total 3,000 Monte Carlo simulations were performed for each scenario. The impact of model misspecification was also explored for the proposed stratified estimator with calibrated weights. Instead of using a logit link in the working model, a cloglog link was used for the misspecified model. 

\subsection{Results}

\subsubsection{Type I Error Study}
\label{typeIerror_causal2}
The results under the correct specification of the PS model are presented in Table \ref{causal_WR_cluster_typeIerror}. The unadjusted estimator ignoring the cluster effect produces large bias and inflated type I error rate, especially when the within-cluster correlation is high. In all the scenarios, the calibration estimator produces the smallest bias. The bias of the calibration estimator tends to be slightly larger under unbalance designs. However, the discrepancy decreases when the number of clusters increases. The random effects estimator shows the largest bias. This is not surprising due to the shrinkage estimation of the random effects in the generalized mixed model. The fixed effects estimator produces larger bias compared to the calibration estimator, and the bias increases when the number of clusters increases. When the number of clusters is large, the fixed effects model tends to be unstable due to the large number of parameters estimated in the model. The bootstrap variance estimation underestimates the variance of the fixed effects and random effects estimators, but provides the variances of the calibration estimator close to the empirical variances. In general, the calibration estimator adequately maintains the nominal type I error level. When the number of clusters is small, the calibration estimator has slightly inflated type I error rate. Overall, under the setting of correct model specification, the fixed and the random effects estimators perform poorly, while the calibration estimator performs well, in terms of bias, variance estimation and type I error controls.  
\\ \\
We further investigated the impacts of different distributional assumptions on calibration estimator only (Table \ref{causal_WR_cluster_typeIerror_mis}). We considered different distributions of cluster effects on treatment selections and the misspecification of the error distribution of the PS model. Estimators with a misspecified link function of the PS model had similar performance with the one with the correctly specified link function. Specifically with the KL distance we chose for the calibration estimator, it produced small bias for both normal and gamma distributions of cluster effects on treatment selections. Type I errors for the gamma distribution were inflated, especially for the small number of clusters, but type I errors were closer to the nominal level when the number of clusters increased. Overall, the calibration estimator with KL distance performs better if one assumes normal distribution for the cluster effects on treatment selections. Nevertheless, the performance improves for other distributions if the number of clusters increases. Here, we argue that generally it's reasonable to assume normal distribution for cluster effects. Additionally, when the number of clusters is large, the distribution of the cluster effects will approach normal. Therefore, the proposed calibration estimator will have reasonable performance in settings where the number of clusters is moderate or large.

%%%%%%%%  Under correct PS model   %%%%%%%%%%
% \usepackage{multirow}
% \usepackage{graphicx}
\begin{table}[!htbp]	
	\centering
	\caption{Type I Error Study on the Causal Win Ratio for Clustered-Dependent Subjects}
	\label{causal_WR_cluster_typeIerror}
	\resizebox{0.7\textwidth}{!}{%
		\begin{threeparttable}
			
			\begin{tabular}{l|cccccc}
				\hline\hline
				\multicolumn{1}{c|}{}              &          & Estimator     & Bias   & Empirical SE & Estimated SE & Type I Error \\ \hline
				\multirow{24}{*}{$m=20$, $n_i=50$} & \multicolumn{6}{l}{\% Treatment Assignment: 50\%}                             \\
				&          &               &        &              &              &              \\
				& ICC=0.07 & Unadjusted    & -0.108 & 0.111        & 0.105        & 0.187        \\
				&          & Logistic      & -0.232 & 0.128        & 0.118        & 0.508        \\
				&          & Fixed  & 0.085  & 0.330        & 0.201        & 0.101        \\
				&          & Random & 0.294  & 0.481        & 0.176        & 0.523        \\
				&          & Calibration   & 0.002  & 0.141        & 0.136        & 0.064        \\ \cline{3-7} 
				& ICC=0.2  & Unadjusted    & -0.184 & 0.145        & 0.106        & 0.426        \\
				&          & Logistic      & -0.233 & 0.129        & 0.120        & 0.497        \\
				&          & Fixed  & 0.085  & 0.334        & 0.204        & 0.098        \\
				&          & Random & 0.293  & 0.493        & 0.179        & 0.529        \\
				&          & Calibration   & 0.000  & 0.143        & 0.139        & 0.065        \\ \cline{2-7} 
				& \multicolumn{6}{l}{\% Treatment Assignment: 30\%}                             \\
				&          &               &        &              &              &              \\
				& ICC=0.07 & Unadjusted    & -0.103 & 0.111        & 0.114        & 0.157        \\
				&          & Logistic      & -0.216 & 0.136        & 0.129        & 0.404        \\
				&          & Fixed  & 0.118  & 0.338        & 0.215        & 0.096        \\
				&          & Random & 0.278  & 0.447        & 0.187        & 0.486        \\
				&          & Calibration   & 0.008  & 0.151        & 0.159        & 0.048        \\ \cline{3-7} 
				& ICC=0.2  & Unadjusted    & -0.179 & 0.143        & 0.114        & 0.379        \\
				&          & Logistic      & -0.216 & 0.137        & 0.129        & 0.395        \\
				&          & Fixed  & 0.119  & 0.338        & 0.216        & 0.094        \\
				&          & Random & 0.275  & 0.451        & 0.189        & 0.479        \\
				&          & Calibration   & 0.010  & 0.153        & 0.161        & 0.047        \\ \hline
				\multirow{24}{*}{$m=50$, $n_i=20$} & \multicolumn{6}{l}{\% Treatment Assignment: 50\%}                             \\
				&          &               &        &              &              &              \\
				& ICC=0.07 & Unadjusted    & -0.109 & 0.096        & 0.104        & 0.165        \\
				&          & Logistic      & -0.223 & 0.123        & 0.121        & 0.454        \\
				&          & Fixed  & 0.202  & 0.373        & 0.247        & 0.108        \\
				&          & Random & 0.366  & 0.582        & 0.193        & 0.567        \\
				&          & Calibration   & 0.001  & 0.129        & 0.133        & 0.044        \\ \cline{3-7} 
				& ICC=0.2  & Unadjusted    & -0.189 & 0.114        & 0.105        & 0.451        \\
				&          & Logistic      & -0.224 & 0.124        & 0.123        & 0.441        \\
				&          & Fixed  & 0.199  & 0.375        & 0.250        & 0.101        \\
				&          & Random & 0.363  & 0.584        & 0.196        & 0.569        \\
				&          & Calibration   & 0.000  & 0.133        & 0.135        & 0.050        \\ \cline{2-7} 
				& \multicolumn{6}{l}{\% Treatment Assignment: 30\%}                             \\
				&          &               &        &              &              &              \\
				& ICC=0.07 & Unadjusted    & -0.109 & 0.095        & 0.113        & 0.120        \\
				&          & Logistic      & -0.208 & 0.131        & 0.132        & 0.369        \\
				&          & Fixed  & 0.236  & 0.412        & 0.269        & 0.137        \\
				&          & Random & 0.355  & 0.550        & 0.212        & 0.508        \\
				&          & Calibration   & 0.000  & 0.148        & 0.150        & 0.044        \\ \cline{3-7} 
				& ICC=0.2  & Unadjusted    & -0.186 & 0.111        & 0.113        & 0.380        \\
				&          & Logistic      & -0.208 & 0.133        & 0.133        & 0.357        \\
				&          & Fixed  & 0.236  & 0.407        & 0.269        & 0.148        \\
				&          & Random & 0.363  & 0.553        & 0.212        & 0.512        \\
				&          & Calibration   & 0.002  & 0.150        & 0.152        & 0.040        \\ \hline
			\end{tabular}%
			\begin{tablenotes}
				\footnotesize
				\item $m$ is the number of clusters and $n_i$ is the cluster size for the $i^{th}$ cluster, where $i=1,...,m$. Equal cluster sizes are assumed across all clusters, however, the number of treatment and control patients within the same cluster could be different. The Unadjusted estimator is the IPW causal win ratio estimator assuming subjects are independent; The Logistic, Fixed effects, Random effects, and the Calibration estimators use the weighted stratified estimator. The Logistic estimator is with the PS estimated from logistic regression without considering cluster effects; the Fixed effects estimator is with the PS estimated from logistic regression with cluster effects as fixed effects; the Random effects estimator is with the PS estimated from logistic regression with cluster effects as random effects; the Calibration estimator is with the PS estimated from calibration method using the correct working model (logit link). 
				\item Acronym: SE: standard error, ICC: intra-cluster correlation, PS: propensity score
			\end{tablenotes}
		\end{threeparttable}
	}
\end{table}

%%%%%%% Investigate Type I error under different distribution assumptions for PS model and distribution of cluter effect for treatment selection %%%%%%%%%%%%%%%%%%%%
% \usepackage{multirow}
% \usepackage{graphicx}
\begin{table}[!htbp]
	\centering
	\caption{Type I Error Study on Calibration Estimator under Model Misspecification}
	\label{causal_WR_cluster_typeIerror_mis}
	\resizebox{0.95\textwidth}{!}{%
		\begin{threeparttable}
			
			\begin{tabular}{l|cccccc}
				\hline\hline
				\multicolumn{1}{c|}{}            &          & Estimator              & Bias   & Empirical SE & Estimated SE & Type I Error \\ \hline
				\multirow{20}{*}{$m=20, n_i=50$} & \multicolumn{6}{l}{\% Treatment Assignment: 50\%}                                       \\
				&          &                        &        &              &              &              \\
				& ICC=0.07 & True\_Normal PS+logit  & 0.002  & 0.141        & 0.136        & 0.058        \\
				&          & Mis\_Normal PS+cloglog & 0.002  & 0.140        & 0.135        & 0.059        \\
				&          & True\_Gamma PS+logit   & 0.004  & 0.095        & 0.091        & 0.077        \\
				&          & Mis\_Gamma PS+cloglog  & 0.004  & 0.095        & 0.090        & 0.079        \\ \cline{3-7} 
				& ICC=0.2  & True\_Normal PS+logit  & 0.000  & 0.143        & 0.139        & 0.061        \\
				&          & Mis\_Normal PS+cloglog & 0.000  & 0.142        & 0.138        & 0.064        \\
				&          & True\_Gamma PS+logit   & 0.004  & 0.096        & 0.092        & 0.078        \\
				&          & Mis\_Gamma PS+cloglog  & 0.004  & 0.095        & 0.092        & 0.077        \\ \cline{2-7} 
				& \multicolumn{6}{l}{\% Treatment Assignment: 30\%}                                       \\
				&          &                        &        &              &              &              \\
				& ICC=0.07 & True\_Normal PS+logit  & 0.008  & 0.151        & 0.159        & 0.043        \\
				&          & Mis\_Normal PS+cloglog & 0.009  & 0.151        & 0.158        & 0.044        \\
				&          & True\_Gamma PS+logit   & 0.001  & 0.114        & 0.108        & 0.069        \\
				&          & Mis\_Gamma PS+cloglog  & 0.002  & 0.114        & 0.108        & 0.069        \\ \cline{3-7} 
				& ICC=0.2  & True\_Normal PS+logit  & 0.010  & 0.153        & 0.161        & 0.041        \\
				&          & Mis\_Normal PS+cloglog & 0.011  & 0.153        & 0.161        & 0.041        \\
				&          & True\_Gamma PS+logit   & 0.002  & 0.116        & 0.110        & 0.063        \\
				&          & Mis\_Gamma PS+cloglog  & 0.003  & 0.115        & 0.110        & 0.061        \\ \hline
				\multirow{20}{*}{$m=50, n_i=20$} & \multicolumn{6}{l}{\% Treatment Assignment: 50\%}                                       \\
				&          &                        &        &              &              &              \\
				& ICC=0.07 & True\_Normal PS+logit  & 0.001  & 0.129        & 0.133        & 0.042        \\
				&          & Mis\_Normal PS+cloglog & 0.001  & 0.129        & 0.133        & 0.041        \\
				&          & True\_Gamma PS+logit   & -0.002 & 0.100        & 0.098        & 0.057        \\
				&          & Mis\_Gamma PS+cloglog  & -0.002 & 0.099        & 0.098        & 0.060        \\ \cline{3-7} 
				& ICC=0.2  & True\_Normal PS+logit  & 0.000  & 0.133        & 0.136        & 0.049        \\
				&          & Mis\_Normal PS+cloglog & 0.000  & 0.132        & 0.135        & 0.047        \\
				&          & True\_Gamma PS+logit   & -0.003 & 0.102        & 0.100        & 0.065        \\
				&          & Mis\_Gamma PS+cloglog  & -0.003 & 0.101        & 0.099        & 0.069        \\ \cline{2-7} 
				& \multicolumn{6}{l}{\% Treatment Assignment: 30\%}                                       \\
				&          &                        &        &              &              &              \\
				& ICC=0.07 & True\_Normal PS+logit  & 0.001  & 0.148        & 0.151        & 0.041        \\
				&          & Mis\_Normal PS+cloglog & 0.002  & 0.148        & 0.150        & 0.041        \\
				&          & True\_Gamma PS+logit   & 0.004  & 0.118        & 0.122        & 0.043        \\
				&          & Mis\_Gamma PS+cloglog  & 0.006  & 0.118        & 0.121        & 0.046        \\ \cline{3-7} 
				& ICC=0.2  & True\_Normal PS+logit  & 0.003  & 0.150        & 0.153        & 0.039        \\
				&          & Mis\_Normal PS+cloglog & 0.004  & 0.150        & 0.153        & 0.041        \\
				&          & True\_Gamma PS+logit   & 0.005  & 0.120        & 0.124        & 0.045        \\
				&          & Mis\_Gamma PS+cloglog  & 0.007  & 0.119        & 0.123        & 0.049        \\ \hline
			\end{tabular}%
			\begin{tablenotes}
				\footnotesize
				\item $m$ is the number of clusters and $n_i$ is the cluster size for the $i^{th}$ cluster, where $i=1,...,m$. Equal cluster sizes are assumed across all clusters, although the number of treatment and control patients within the same cluster could be different. ``Normal PS"/``Gamma PS" indicates that data are generated using normal/gamma distribution for the cluster effects on treatment selections. The treatment status are generated using logit link for all scenarios. ``logit"/``cloglog" indicates that the working PS model is fitted using logit/cloglog link. ``True\_"/``Mis\_" indicates that the working PS model is correctly/incorrectly specified.   
				\item Acronym: SE: standard error, ICC: intra-cluster correlation, PS: propensity score
			\end{tablenotes}
		\end{threeparttable}
		
	}
\end{table}

\subsubsection{Power Analysis}
\begin{figure}[!htbp]
	\begin{threeparttable}
		
		\begin{center}
			
			\includegraphics[width=\textwidth]{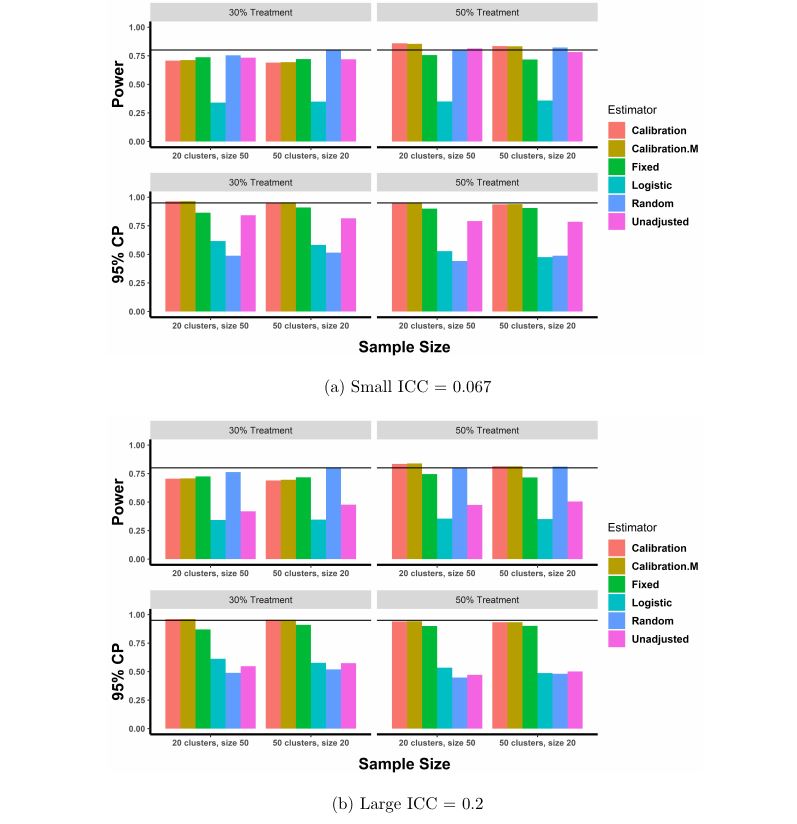}

		\end{center}
		\begin{tablenotes}
			\scriptsize
			\item The order of the estimators in the legend from top to bottom corresponds to the bars from left to right in the plot panels. The horizontal reference lines in the power plots are at 0.8. The horizontal reference lines in the coverage probability plots are at 0.95. The Unadjusted estimator is the IPW causal win ratio estimator assuming subjects are independent; The Logistic, Fixed effects, Random effects, and the Calibration estimators use the weighted stratified estimator. The Logistic estimator is with the PS estimated from logistic regression without considering cluster effects; the Fixed effects estimator is with the PS estimated from logistic regression with cluster effects as fixed effects; the Random effects estimator is with the PS estimated from logistic regression with cluster effects as random effects; the Calibration estimator is with the PS estimated from calibration method using the correct working model (logit link); the Calibration.M estimator is with the PS estimated from calibration method using the incorrect working model (cloglog link).
			\item Acronym: CP: coverage probability, PS: propensity score
		\end{tablenotes}
	\end{threeparttable}
	\caption{%
		Power Analysis of Causal Win Ratio for Dependent Subjects
	}%
	\label{power_depPS}
\end{figure}

Power analysis was conducted under the alternative hypothesis $\log(\mu)=0.42$. The power and the 95\% coverage probability for the true $\log(\mu)$ for all estimators were evaluated using the same simulation settings (Figure \ref{power_depPS}). All estimators have similar or slightly less power in the large ICC setting compared to the small ICC setting. Generally, when there is equal percentage of treatment assignments, the power is higher than the one with unequal percentage of treatment assignments. The calibration estimators with and without correctly specified working model have similar powers in all scenarios, and both of the estimators have close to 95\% coverage of the true $\log(\mu)$. In cases where the fixed and random effects estimators have higher power than the calibration estimators, the coverage probability of fixed and random effects estimators are much less than 95\% due to the inflated type I errors shown in Table \ref{causal_WR_cluster_typeIerror}.

%% power and CP for small vs. large ICC

%%%%%%%%%%%%%%%%%%%%%%%%%%%%%%%%%%%%%%%%%%%%%%%%%%%%%%%%%%%%%%%%%%%%%%%%%%%%%%%%%%%%%%%%%%%%%%%%%%%%%%%%%%%%%%%%%%%%%%%%%%%%%%%%%%%%%%%%%%%%%%%%%%
%%%%%%%%%%%%%%%%%%%%%%%%%%%%%%%%%%%%%%%%%%%%%%%%%%%%%%%%%%%%%%%%%%%%%%%%%%%%%%%%%%%%%%%%%%%%%%%%%%%%%%%%%%%%%%%%%%%%%%%%%%%%%%%%%%%%%%%%%%%%%%%%%%
%%%%%%%%%%%%%%%%%%%%%%%%%%%%%%%%%%%%%%%%%%%%%%%%%%%%%%%%%%%%%%%%%%%%%%%%%%%%%%%%%%%%%%%%%%%%%%%%%%%%%%%%%%%%%%%%%%%%%%%%%%%%%%%%%%%%%%%%%%%%%%%%%%
%%%%%%%%%%%%%%%%%%%%%%%%%%%%%%%%%%%%%%%%%%%%%%%%%%%%%%%%%%%%%%%%%%%%%%%%%%%%%%%%%%%%%%%%%%%%%%%%%%%%%%%%%%%%%%%%%%%%%%%%%%%%%%%%%%%%%%%%%%%%%%%%%%

\vspace{-3mm}
\section{Data Example}
\begin{figure}[!htbp]
	\begin{center}

		\includegraphics[width=\textwidth]{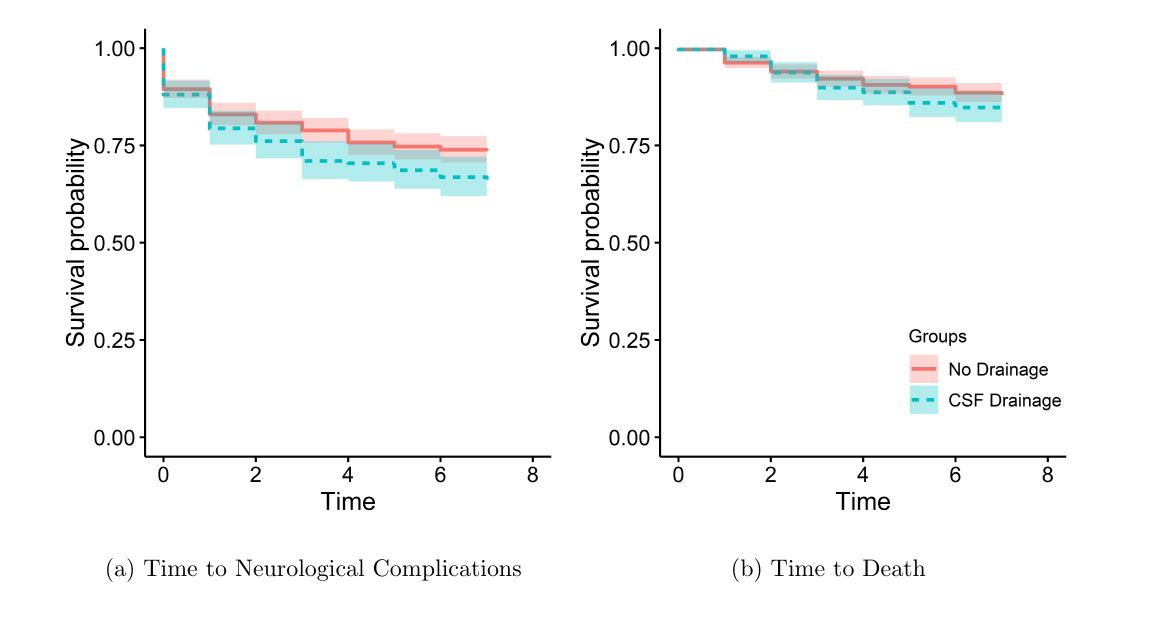}

	\end{center}
	\caption{Kaplan-Meier Curves of Time to Events in First 7 Days of Hospitalization}
	\label{time_neuro_death}
\end{figure}

\vspace{-3mm}
\label{calibrationWR_example}
Traumatic brain injury (TBI) remains the leading cause of death and disability in children. Despite the preventative measures, thousands of children every year in the US and abroad are affected \citep{michaud1993traumatic, kurihara2000traumatic, mckinlay2008prevalence}. None of the 33 randomized controlled trials (RCTs) of various therapies so far has demonstrated the improved patients outcomes. The Approaches and Decisions for Acute Pediatric TBI (ADAPT) trial was an observational study to investigate the impacts of various medical interventions on children under 18 years old with TBI. For demonstrative purposes, we focus on the comparison between the Cerebrospinal Fluid (CSF) drainage intervention and no CSF treatment in the first 7 days of hospitalization after brain injuries. Total 880 patients from 37 sites were included in this analysis. Thirty-four percent (34\%) of the patients were treated with CSF and 66\% of them had no CSF treatment. The sites are clusters and we assume that patients in the same site share similar cluster level characteristics. For instance, medical facilities, resources and physician training may be different across sites, but similar within sites. All sites have patients from both groups. We are interested in the composite endpoints of time to first most severe neurological complication and time to death. Intuitively, we value death as a more important event than complications, and we want to reflect this information in the analysis. Total 26 confounders were considered, including patient demographics, TBI causes and several pre-hospitalization measures. Missing values of confounders were imputed using multiple imputation implemented in the R package \verb|mice| \citep{buuren2010mice} with default settings.  
\\ \\
All patients were followed up daily during hospitalization. Total 13\% of the patients died and 29\% had neurological complications. Nine percent (9\%) of them experienced neurological complications before their death, and 4\% died without complications. About 20\% of the patients experienced neurological complications but survived the first 7 days. The Kaplan-Meier curves of the time to neurological complication and time to death between groups are presented in Figure \ref{time_neuro_death}. For both endpoints, the no drainage group tends to be more beneficial compared to the CSF drainage group. The difference between groups is bigger in time to complications than in time to death. 
\\ \\
We first compared the causal win ratio approach to traditional survival analysis methods accounting for confounders, assuming subjects are independent (Table \ref{causal_timeFirst_WR}). The traditional methods include analyses on time-to-complications only, time-to-death only, and time-to-first-event analysis for the composite endpoints. We considered both Cox proportional hazards model with IPWs and additive hazards model with IPWs \citep{lin1994semiparametric}. \cite{aalen2015does} noted that the hazard ratio in a Cox model is not a natural causal quantity to consider, thereby we also provided analysis using additive hazards model for the reader to consider. The CSF drainage group was the treatment of interest and the group without CSF treatment was the reference group. The probability of receiving the CSF treatment conditional on pre-specified confounders (or PS) was calculated using a logistic regression model. The IPWs were calculated as the reciprocal of the PS for CSF drainage group and the reciprocal of one minus the PS for the reference group. Under the Cox model, the treatment effect estimate is the hazard ratio (HR) on a log scale, and a positive value indicates that the reference group is favored. Under the additive hazards model, the effect estimate is the absolute difference of the hazards, and a positive value indicates that the reference group is favored. The effect estimate from the causal win ratio is the win ratio (WR) on a log scale, and a negative value indicates that the reference group is beneficial. For single-event analyses, the hazard of neurological complications or death for CSF drainage patients is slightly higher than the one for the no drainage patients. Since the time-to-first-event analysis treats
both endpoints equally important and complications always happen before death, the neurological complications drive the results of the traditional time-to-first-event analysis. For the win ratio approach, the IPW estimator was implemented according to Equation (\ref{ipw}) in Section \ref{wr_independent_confounding}. We observe that being in the CSF group is 15\% less beneficial than in the no drainage group in delaying time-to-neurological-complications at least, if not time-to-death as the primary event, yet the difference is insignificant.

%%%  Kaplan-Meier curves for time to events for ADAPT trial

%%%  Compare time to first event analysis to causal win ratio, account for confounders but not cluster effects. 
% \usepackage{multirow}
% \usepackage{graphicx}
\begin{table}[!htbp]
	\caption{Causal Inference of Time-to-First-Event Analysis and Win Ratio for Independent Subjects}
	\label{causal_timeFirst_WR}
	\centering
	\resizebox{\textwidth}{!}{%
		\begin{threeparttable}
			\begin{tabular}{c|ccccccc}
				\hline
				\textbf{Method}                  & \textbf{Endpoint}                        & \textbf{$\beta$}                 & \textbf{$\exp(\beta)$}             & \textbf{$se(\beta)$}            & \textbf{95\% CI}                       & \textbf{Test Statistic}          & \textbf{P-value}                \\ \hline
				Additive Hazards with IPW   & Death Only      & 0.002    &     -      & 0.004 & (-0.0058, 0.0099) & 0.513       & 0.608   \\
				($\beta:$ difference in hazards) &  Neurological Complications Only     & 0.009    &      -     & 0.007 & (-0.0049, 0.0237) & 1.292       & 0.196   \\
				& Composite (Time-to-First-Event)      & 0.011    &     -      & 0.008 & (-0.0047, 0.0268) & 1.377       & 0.168   \\ \hline
				Cox Regression with IPW & Death Only                      & 0.097                   & 1.102                  & 0.188                  & (-0.27, 0.46)                 & 0.516                   & 0.606                  \\
				($\beta: \log(HR)$)      & Neurological Complications Only & 0.169                   & 1.184                  & 0.125                  & (-0.08, 0.41)                 & 1.353                   & 0.176                  \\
				& Composite (Time-to-First-Event) & 0.170                   & 1.185                  & 0.117                  & (-0.06, 0.4)                  & 1.453                   & 0.146                  \\ \hline
				Causal Win Ratio        & \multirow{2}{*}{Composite}      & \multirow{2}{*}{-0.158} & \multirow{2}{*}{0.854} & \multirow{2}{*}{0.125} & \multirow{2}{*}{(-0.4, 0.09)} & \multirow{2}{*}{-1.264} & \multirow{2}{*}{0.206} \\
				($\beta: \log(WR)$)      &                                 &                         &                        &                        &                               &                         &                        \\ \hline
			\end{tabular}%
			\begin{tablenotes}
				\footnotesize
				\item $\beta$ is $\log(HR)$ for Cox regression and $\log(WR)$ for win ratio approach. 95\% CI is for $\beta$.
				\item Acronym: IPW: inverse probability weights, HR: hazard ratio, WR: win ratio, se: standard error, CI: confidence interval   
			\end{tablenotes}
		\end{threeparttable}
		
	}
\end{table}

Second, we compared the three weighted stratified causal WR estimators discussed in the simulation study, accounting for cluster effects. For the calibration esitmator, the initial weights were first calculated using IPWs, and the calibrated weights were estimated according to Section \ref{weight_estimation}. Then the cluster-specific weighted win ratios were estimated based on Equation (\ref{cluster_tao}), and the estimated average treatment effect of the causal win ratio for clustered data was estimated acccording to Equation (\ref{equ7}). The balance of covariates between groups in overall sample was calculated using the absolute mean difference criterion \citep{austin2015moving}. As shown in Figure 3, the distributions of covariates between groups achieve exact balance for the calibration estimator, while imbalance are observed for some covariates when using the weights calculated by the logistic regression, fixed or random effects models. Without accounting for cluster effects, the unadjusted estimator yields a larger effect estimate compared to other cluster-adjusted weighted estimators (Table \ref{data_multipleEst}). Although the logistic estimator does not account for cluster effects in the treatment selection process, i.e. estimation of weights, the effect estimate by the logistic estimator is close to the one from the calibration estimator. This implies that, on average, the treatment decision processes were similar by physicians across sites. However, by taking the ratio between the variances of the two $log(WR)$, the calibration estimator shows 37\% more variability of the treatment effects compared to logistic estimator. Comparing the logistic estimator with the unadjusted estimator, the logistic estimator has 52\% more variation in the treatment effect than the unadjusted estimator. The fixed effects and random effects estimators have the effect estimates with opposite signs compared to the other estimators, but all of their 95\% confidence intervals include 0, leading to the same conclusions in hypothesis testing. Recall the simulation results in Section \ref{typeIerror_causal2}, the fixed effects and random effects estimators have much greater variations compared to the calibration estimator, but the SEs are similar among these three estimators in this real example. This may be due to the large number of confounders included in the analysis. Finally, we interpret the result based on the calibration weighted estimator as follows. Accounting for the potential confounders and cluster effects, being in the CSF group is about 3\% less beneficial than in the no drainage group in delaying time-to-neurological-complications at least, if not time-to-death as the primary event, yet the difference is insignificant. There is no sufficient evidence to support any causal relationship between CSF drainage treatment and time-to-neurological-complication or time-to-death during the 7-day follow-up period.

\begin{figure}[!htbp]
	\begin{threeparttable}
		\begin{center}
			\includegraphics[width=0.8\textwidth]{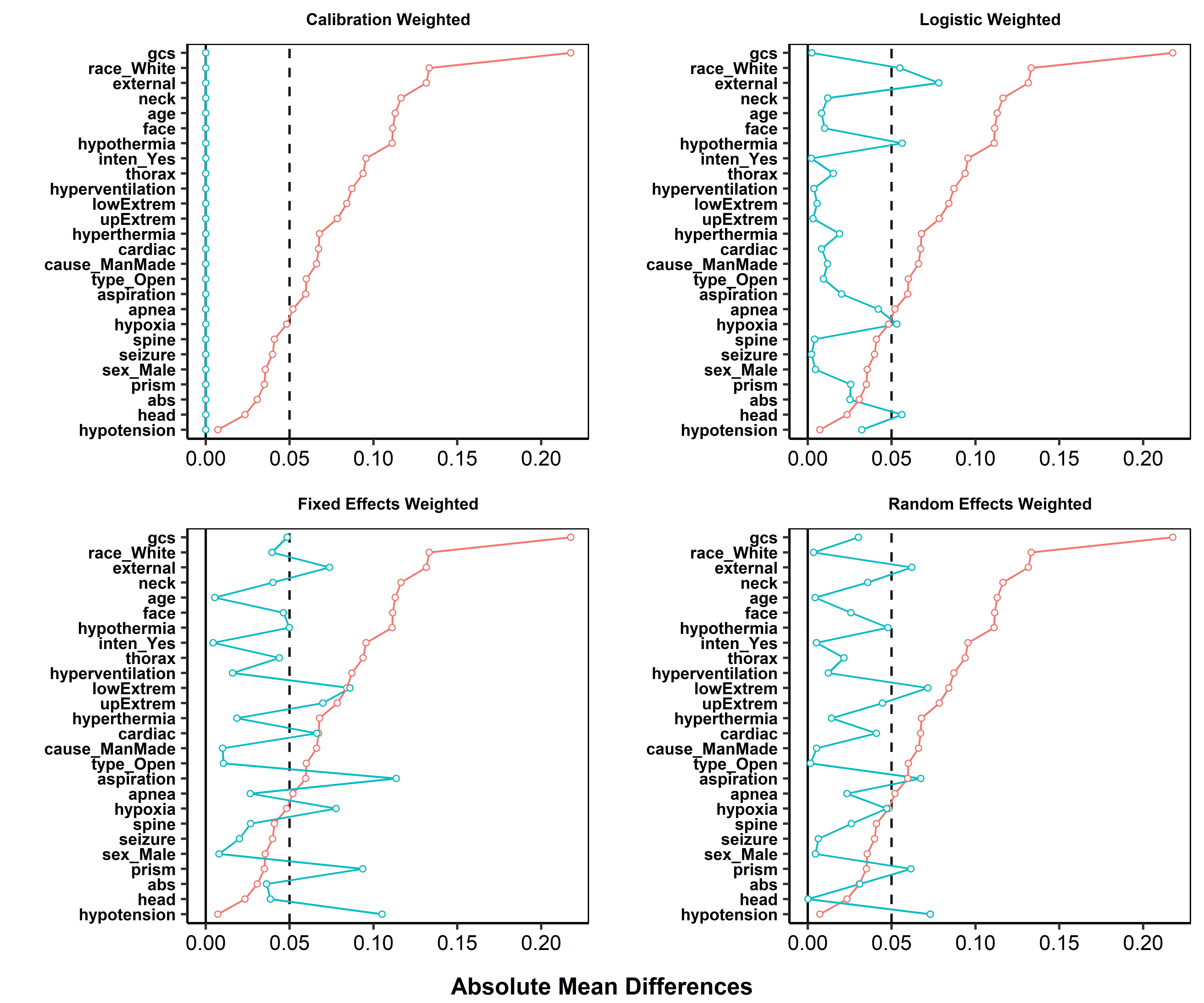}
		\end{center}
		\begin{tablenotes}
			\footnotesize
			\item Blue line indicates the balance between groups for cluster-adjusted weighted estimators, and the red line indicates the balance between groups for the unadjusted estimator. Smaller absolute mean difference implies better balance between groups. The vertical black solid reference line is at absolute mean difference = 0. The vertical black dash reference line is at absolute mean difference = 0.05. 
		\end{tablenotes}
	\end{threeparttable}
	\caption{Balance of Covariates between Groups in Overall Sample}
	\label{balance}
\end{figure}

% \usepackage{graphicx}
\begin{table}[!htbp]
	\caption{Data Analysis using Cluster-Adjusted Weighted Estimators}
	\label{data_multipleEst}
	\centering
	\resizebox{0.8\textwidth}{!}{%
		\begin{threeparttable}
			\begin{tabular}{c|cccccc}
				\hline
				\textbf{Estimator} & \textbf{$\beta$} & \textbf{$\exp(\beta)$} & \textbf{$se(\beta)$} & \textbf{95\% CI} & \textbf{Test Statistic} & \textbf{P-value} \\ \hline
				Unadjusted         & -0.158           & 0.854                 & 0.125                & (-0.4, 0.09)     & -1.264                  & 0.206            \\ \hline
				Logistic           & -0.024           & 0.977                 & 0.154                & (-0.32, 0.28)    & -0.154                  & 0.878            \\ \hline
				Fixed Effects      & 0.062            & 1.064                 & 0.211                & (-0.35, 0.48)    & 0.295                   & 0.768            \\ \hline
				Random Effects     & 0.006            & 1.006                 & 0.176                & (-0.34, 0.35)    & 0.036                   & 0.971            \\ \hline
				Calibration        & -0.034           & 0.966                 & 0.180                & (-0.39, 0.32)    & -0.191                  & 0.848            \\ \hline
			\end{tabular}%
			\begin{tablenotes}
				\footnotesize
				\item $\beta$ is $\log(WR)$. 95\% CI is for $\beta$.
				\item Acronym: WR: win ratio, se: standard error, CI: confidence interval   
			\end{tablenotes}
		\end{threeparttable}
	}
\end{table}

%%%%%%%%%%%%%%%%%%%%%%%%%%%%%%%%%%%%%%%%%%%%%%%%%%%%%%%%%%%%%%%%%%%%%%%%%%%%%%%%%%%%%%%%%%%%%%%%%%%%%%%%%%%%%%%%%%%%%%%%%%%%%%%%%%%%%%%%%%%%%%%%%%
%%%%%%%%%%%%%%%%%%%%%%%%%%%%%%%%%%%%%%%%%%%%%%%%%%%%%%%%%%%%%%%%%%%%%%%%%%%%%%%%%%%%%%%%%%%%%%%%%%%%%%%%%%%%%%%%%%%%%%%%%%%%%%%%%%%%%%%%%%%%%%%%%%
%%%%%%%%%%%%%%%%%%%%%%%%%%%%%%%%%%%%%%%%%%%%%%%%%%%%%%%%%%%%%%%%%%%%%%%%%%%%%%%%%%%%%%%%%%%%%%%%%%%%%%%%%%%%%%%%%%%%%%%%%%%%%%%%%%%%%%%%%%%%%%%%%%
%%%%%%%%%%%%%%%%%%%%%%%%%%%%%%%%%%%%%%%%%%%%%%%%%%%%%%%%%%%%%%%%%%%%%%%%%%%%%%%%%%%%%%%%%%%%%%%%%%%%%%%%%%%%%%%%%%%%%%%%%%%%%%%%%%%%%%%%%%%%%%%%%%

\section{Discussions}
\label{calibrationWR_conclusion}
Composite endpoints are commonly used in clinical trials and observational studies with an anticipation that clinically relevant endpoints as a whole would yield meaningful treatment benefits \citep{cordoba2010definition, capodanno2016computing, weintraub2016statistical}. For some diseases such as cardiovascular disease, the composite endpoints are often required to be the primary endpoint for clinical studies \citep{tong2012weighting}. When the composite endpoints imply different clinical importance, the analysis should take into account the magnitude of importance of each endpoint to reach a more sensible conclusion. For example, progression-free survival is a commonly used endpoint in cancer studies with death as fatal (or terminal) event and disease progression as non-fatal event. Death is more important than disease progression. However, the traditional time-to-first-event analysis emphasizes disease progression since it always happens before death. Therefore, accounting for the importance of endpoints in a hierarchical fashion would provide clinically valuable interpretation of the data.
\\
A calibration weighted stratified win ratio estimator was proposed for causal inference for the composite endpoints with cluster-dependent subjects. This estimator is able to prioritize the important endpoint, and account for confounders and correlations within clusters. The calibration weights create balanced covariates and cluster effects distributions between groups. Additionally, they are robust against the distribution assumptions of the treatment selection modeling. Compared to fixed or random effects estimator, the proposed estimator showed superior performances in terms of bias, variance, type I error and power, regardless of the percentage of treatment assignments and intra-cluster correlations. We focus on two-group comparison in our current study; however, the proposed approach can be easily extended to settings with multiple-treatment comparisons. A multinomial logistic regression can be used to calculate the initial weights for the calibration weight estimation and similar procedures can be applied to estimate the win ratio as treatment effect. Additionally, the proposed calibration weighted stratified win ratio estimator can be further extended to hierarchical structures that have more than two levels, which might merit further investigation.

%%%%%%%%%%%%%%%%%%%%%%%%%%%%%%%%%%%%%%%%%%%%%%%%%%%%%%%%%%%%%%%%%%%%%%%%%%%%%%%%%%%%%%%%%%%%%%%%%%%%%%%%%%%%%%%%%%%%%%%%%%%%%%%%%%%%%%%%%%%%%%%%%%
%%%%%%%%%%%%%%%%%%%%%%%%%%%%%%%%%%%%%%%%%%%%%%%%%%%%%%%%%%%%%%%%%%%%%%%%%%%%%%%%%%%%%%%%%%%%%%%%%%%%%%%%%%%%%%%%%%%%%%%%%%%%%%%%%%%%%%%%%%%%%%%%%%
%%%%%%%%%%%%%%%%%%%%%%%%%%%%%%%%%%%%%%%%%%%%%%%%%%%%%%%%%%%%%%%%%%%%%%%%%%%%%%%%%%%%%%%%%%%%%%%%%%%%%%%%%%%%%%%%%%%%%%%%%%%%%%%%%%%%%%%%%%%%%%%%%%
%%%%%%%%%%%%%%%%%%%%%%%%%%%%%%%%%%%%%%%%%%%%%%%%%%%%%%%%%%%%%%%%%%%%%%%%%%%%%%%%%%%%%%%%%%%%%%%%%%%%%%%%%%%%%%%%%%%%%%%%%%%%%%%%%%%%%%%%%%%%%%%%%%

%\backmatter

%\section*{Acknowledgments}

%\subsection*{Author contributions}

%\subsection*{Financial disclosure}

%None reported.

%\subsection*{Disclaimer}
%Di Zhang contributed to this article in her personal capacity. The views expressed are her own, and do not necessarily represent the views of the Food and Drug Administration or the United States Government.

%\subsection*{Conflict of interest}

%The authors declare no potential conflict of interests.

%\section*{Supporting Information}

\section*{Appendix 1: Estimation of calibrated weights proposed by Yang \cite{yang2018propensity}.}

\subsection*{Appendix 1.1: Balance of Covariates for Clustered Data}

Based on the features of the propensity score as a balancing score for independent subjects \citep{chan2016globally, imai2014covariate}, for each covariate $X_{ijl}$, we define the propensity score as a balancing score for clustered data as
\begin{equation}
\label{cluster_balance}
E[\frac{Z_{ij} X_{ijl}}{\pi(X_{ij}; \alpha, \gamma_i)}]=E[\frac{(1-Z_{ij})X_{ijl}}{1-\pi(X_{ij}; \alpha, \gamma_i)}]=E[X_{ijl}], 
\end{equation}
where $l=1,...,p$, and $\pi(X_{ij}; \alpha, \gamma_i)$ be the propensity score with parameter $\alpha$, which is the probability of receiving treatment, given covariates and cluster effect. If the propensity score is correct, then the weighted covariates in the treatment group will have similar distributions as the weighted covariates in the control group.
\\ \\
The empirical version of Equation (\ref{cluster_balance}) is
\begin{equation}
\label{cluster_balance_emp1}
\sum_{i=1}^{m} \sum_{j=1}^{n_i} \frac{Z_{ij} X_{ijl}}{\hat{\pi}(X_{ij}; \hat{\alpha}, \hat{\gamma_i})}
=\sum_{i=1}^{m} \sum_{j=1}^{n_i} \frac{(1-Z_{ij}) X_{ijl}}{1-\hat{\pi}(X_{ij}; \hat{\alpha}, \hat{\gamma_i})}
=\sum_{i=1}^{m} \sum_{j=1}^{n_i} X_{ijl}.
\end{equation}
To ensure the balance of the cluster effects between treatment and control group, we want to satisfy additional conditions \citep{yang2018propensity}
\begin{equation}
\label{cluster_balance_gamma}
E[\frac{Z_{ij} \gamma_i}{\pi(X_{ij}; \alpha, \gamma_i)}]=E[\frac{(1-Z_{ij})\gamma_i}{1-\pi(X_{ij}; \alpha, \gamma_i)}]=E[\gamma_i].
\end{equation}
Therefore, given a specific cluster effect $\gamma_i$, Equation (\ref{cluster_balance_gamma}) becomes
\begin{equation}
\label{cluster_balance_condition}
E[\frac{Z_{ij}}{\pi(X_{ij}; \alpha, \gamma_i)}|\gamma_i]=E[\frac{(1-Z_{ij})}{1-\pi(X_{ij}; \alpha, \gamma_i)}|\gamma_i]=E[1|\gamma_i].
\end{equation}
The empirical version of the Equation (\ref{cluster_balance_condition}) for a specific cluster $i$ is 
\begin{equation}
\label{cluster_balance_emp2}
\sum_{j=1}^{n_i} \frac{Z_{ij}}{\hat{\pi}(X_{ij}; \hat{\alpha}, \hat{\gamma_i})}
=\sum_{j=1}^{n_i} \frac{(1-Z_{ij})}{1-\hat{\pi}(X_{ij}; \hat{\alpha}, \hat{\gamma_i})}
=\sum_{j=1}^{n_i}1 = n_i.
\end{equation}
These two constrains (Equation (\ref{cluster_balance_emp1}) and (\ref{cluster_balance_emp2})) are used in the calibration estimation of the weights. They ensure the balance of covariates and cluster effects between groups so that valid causal inference can be made for observational clustered data. 

\subsection*{Appendix 1.2: Estimation}

The procedure of estimating weights using calibration is 
\begin{enumerate}
	\item Estimate the initial weights $d_{ij}$ using some parametric working model, such as logistic regression. Define $d_{ij}=\frac{Z_{ij}}{\pi(X_{ij}; \alpha)}+\frac{1-Z_{ij}}{1-\pi(X_{ij}; \alpha)}$ as the initial weights.  
	\item Estimate the calibration weights $w_{ij}$ using some distance function $D(w_{ij}, d_{ij})$ with constraints. Define $w_{ij}$ as the final weights.
\end{enumerate}
After estimation of the initial weights $\hat{d_{ij}}$, we want to minimize the objective function with multiple constraints, respect to $w_{ij}$:
\begin{equation}
\label{app_objective_fn}
\begin{split}
O=& \sum_{i=1}^{m} \sum_{j=1}^{n_i} w_{ij}\ln \frac{w_{ij}}{d_{ij}}  \\
&+{\lambda_1}^T\{\sum_{i=1}^{m} \sum_{j=1}^{n_i} [Z_{ij} w_{ij} X_{ij}  - X_{ij}]  \}\\
&+{\lambda_2}^T\{\sum_{i=1}^{m} \sum_{j=1}^{n_i} [(1-Z_{ij}) w_{ij} X_{ij} - X_{ij}]  \}\\
=&\sum_{i=1}^{m} \sum_{j=1}^{n_i} w_{ij}\ln \frac{w_{ij}}{d_{ij}}+{\lambda_1}^T\{\sum_{i=1}^{m} \sum_{j=1}^{n_i}C_{ij1}\}+{\lambda_2}^T\{\sum_{i=1}^{m} \sum_{j=1}^{n_i}C_{ij2}\},
\end{split}
\end{equation}
with additional constraints within each cluster $i$,
\begin{equation}
\label{app_objective_fn_add}
\sum_{j=1}^{n_i} Z_{ij} w_{ij}=\sum_{j=1}^{n_i} (1-Z_{ij}) w_{ij}= n_i,
\end{equation}
Let the first derivative of the objective function equal to 0:
\begin{equation}
\begin{split}
&\frac{\partial g}{\partial w_{ij}}=\ln(\frac{w_{ij}}{d_{ij}})+1+{\lambda_1}^T Z_{ij} X_{ij} + {\lambda_2}^T (1-Z_{ij}) X_{ij} =0\\
\Longrightarrow &\hat{w}_{ij}=d_{ij}Z_{ij}\exp(-1-{\lambda_1}^T Z_{ij} X_{ij}) + d_{ij}(1-Z_{ij}) \exp(-1-{\lambda_2}^T (1-Z_{ij}) X_{ij}).
\end{split}
\end{equation}
Since when $Z_{ij}=1$, $Z_{ij}w_{ij}=d_{ij}Z_{ij}\exp(-1-{\lambda_1}^T Z_{ij} X_{ij})$. According to Equation \ref{app_objective_fn_add}, we have
\begin{equation}
\label{app_add_cond1}
\begin{split}
\sum_{j=1}^{n_i} Z_{ij} w_{ij}&= n_i\\
&=\sum_{j=1}^{n_i} d_{ij}Z_{ij}\exp(-1-{\lambda_1}^T Z_{ij} X_{ij}),\\
\Longrightarrow n_i &= \sum_{j=1}^{n_i} d_{ij}Z_{ij}\exp(-1-{\lambda_1}^T Z_{ij} X_{ij}).
\end{split}
\end{equation}
Same reasoning applies to when $1-Z_{ij}=1$,
\begin{equation}
\label{app_add_cond2}
n_i = \sum_{j=1}^{n_i} d_{ij}(1-Z_{ij})\exp(-1-{\lambda_2}^T (1-Z_{ij}) X_{ij}).
\end{equation}
Therefore, we can rewrite the estimator $\hat{w}_{ij}$ with conditions \ref{app_add_cond1} and \ref{app_add_cond2} as
\begin{equation}
\hat{w}_{ij}=\frac{n_i d_{ij}Z_{ij}\exp(-{\lambda_1}^T Z_{ij} X_{ij})}{\sum_{j=1}^{n_i} d_{ij}Z_{ij}\exp(-{\lambda_1}^T Z_{ij} X_{ij})}
+\frac{n_i d_{ij}(1-Z_{ij})\exp(-{\lambda_2}^T (1-Z_{ij}) X_{ij})}{\sum_{j=1}^{n_i} d_{ij}(1-Z_{ij})\exp(-{\lambda_2}^T (1-Z_{ij}) X_{ij})}
\end{equation} 
Since $\hat{w}_{ij}$ is a function of $\hat{\lambda}_1$ and $\hat{\lambda}_2$, we can estimate $\lambda_1$ and $\lambda_2$ iteratively from $\sum_{i=1}^{m} \sum_{j=1}^{n_i}C_{ij1}=0$ and $\sum_{i=1}^{m} \sum_{j=1}^{n_i}C_{ij2}=0$. \\
Different from the initial weights calculated from the propensity score estimates, the calibration weight is calculated directly by mapping the initial weights to the final weights based on a distance function.

\section*{Appendix 2: Empirical Distribution of the Proposed Estimator}
We show the empirical distribution of the proposed weighted stratified causal win ratio estimator with calibrated weight, based on the simulation set-up (Section \ref{set-up}) for ICC=0.067, varying number of clusters and cluster sizes from 10, 20 to 50. The true underlying causal win ratio on log scale is 0. Similar patterns were observed for ICC=0.2.\\
\begin{figure}[!htbp]
	\begin{threeparttable}
		\begin{center}
			\includegraphics[width=0.7\textwidth]{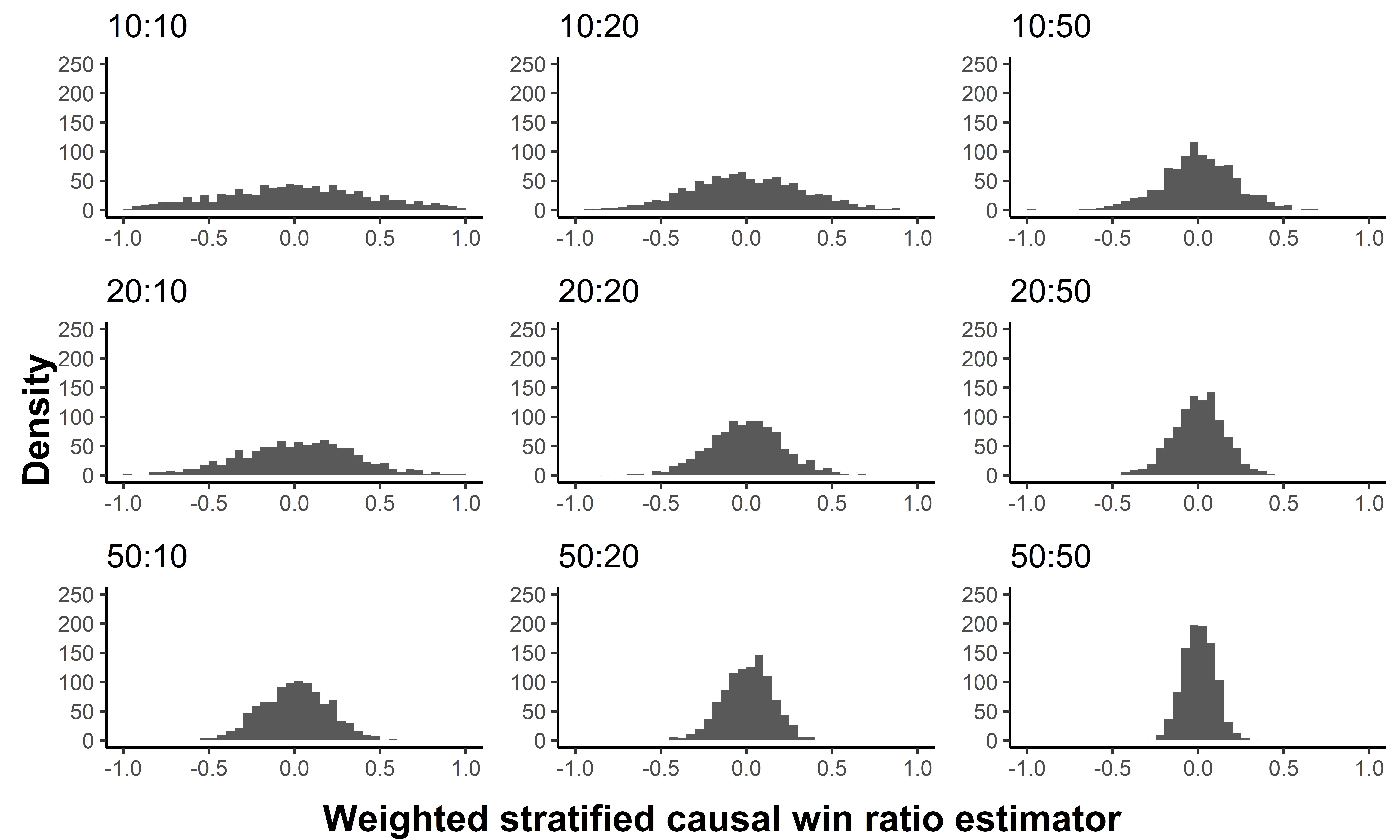}
		\end{center}
		\begin{tablenotes}
			\footnotesize
			\item The title of the inidividual plot indicates "number of clusters : cluster size"
		\end{tablenotes}
	\end{threeparttable}
	\caption{Empirical Distribution of the Proposed Weighted Stratified Estimator with Calibrated Weight on Log Scale}
	\label{balance}
\end{figure}

\newpage
%\nocite{*}% Show all bib entries - both cited and uncited; comment this line to view only cited bib entries;

%\bibliography{refs}%

\begin{thebibliography}{}

\bibitem[{Aalen {\it et al.}(2015)}]{aalen2015does}
Aalen OO, Cook RJ, R{\o}ysland K. (2015). Does Cox analysis of a randomized survival study yield a causal treatment effect? Lifetime Data Analysis 21:4, 579--593.

\bibitem[{Arpino and Cannas(2016)}]{arpino2016propensity}
Arpino B, Cannas M. (2016). Propensity score matching with clustered data: An application to the estimation of the impact of caesarean section on the Apgar score. Statistics in Medicine 35:12, 2074--2091.

\bibitem[{Austin and Stuart(2015)}]{austin2015moving}
Austin PC, Stuart EA. (2015). Moving towards best practice when using inverse probability of treatment weighting (IPTW) using the propensity score to estimate causal treatment effects in observational studies. Statistics in Medicine 34:28, 3661--3679.

\bibitem[{Bebu and Lachin(2015)}]{bebu2015large}
  \bibinfo{author}{Bebu I, Lachin JM}. \bibinfo{year}{(2015)}.
  \bibinfo{title}{Large sample inference for a win ratio analysis of a composite outcome based on prioritized components}. \bibinfo{journal}{Biostatistics} \bibinfo{volume}{17}:\bibinfo{number}{1}, \bibinfo{pages}{178--187}.

\bibitem[{Buuren and Groothuis-Oudshoorn(2010)}]{buuren2010mice}
Buuren S van, Groothuis-Oudshoorn K. (2011).  mice: Multivariate imputation by chained equations in R. Journal of Statistical Software 45:3, 1--68.

\bibitem[{Caillat {\it et al.}(2009)}]{caillat2009package}
Caillat A-L, Dutang C, Dutang MC. (2009). R Package ‘gumbel’.

\bibitem[{Capodanno {\it et al.}(2016)}]{capodanno2016computing}
Capodanno D, Gargiulo G, Buccheri S, Chieffo A, Meliga E, Latib A, Park S-J, Onuma Y, Capranzano P,Valgimigli M {\it et al.} (2016). Computing methods for composite clinical endpoints in unprotected left main coronary artery revascularization: a post hoc analysis of the DELTA registry. JACC: Cardiovascular Interventions 9:22, 2280--2288.

\bibitem[{Chan {\it et al.}(2016)}]{chan2016globally}
Chan KCG, Yam SCP, Zhang Z. (2016).  Globally efficient non-parametric inference of average treatment effects by empirical balancing calibration weighting. Journal of the Royal Statistical Society: Series B (Statistical Methodology) 78:3, 673--700.

\bibitem[{Chen {\it et al.}(2002)}]{chen2002using}
Chen J, Sitter RR, Wu C. (2002). Using empirical likelihood methods to obtain range restricted weights in regression estimators for surveys. Biometrika 89:1, 230--237.

\bibitem[{Cordoba {\it et al.}(2010)}]{cordoba2010definition}
Cordoba G, Schwartz L, Woloshin, Steven and Bae, Harold and G{\o}tzsche, Peter C. (2010). Definition, reporting, and interpretation of composite outcomes in clinical trials: systematic review. British Medical Journal 341, c3920.

\bibitem[{Dong {\it et al.}(2019)}]{dong2019win}
Dong G, Hoaglin DC, Qiu J, Matsouaka RA, Chang Y, Wang J, Vandemeulebroecke M. (2020). The win ratio: On interpretation and handling of ties. Statistics in Biopharmaceutical Research 12:1, 99-106.

\bibitem[{Dong {\it et al.}(2016)}]{dong2016generalized}
Dong G, Li D, Ballerstedt S, Vandemeulebroecke M. (2016). A generalized analytic solution to the win ratio to analyze a composite endpoint considering the clinical importance order among components. Pharmaceutical Statistics 15:5, 430--437.

\bibitem[{Dong {\it et al.}(2018)}]{dong2018stratified}
Dong G, Qiu J, Wang D, Vandemeulebroecke M. (2018). The stratified win ratio. Journal of Biopharmaceutical Statistics 28:4, 778--796.

\bibitem[{Finkelstein and Schoenfeld(2019)}]{finkelstein2019graphing}
Finkelstein DM, Schoenfeld DA. (2019). Graphing the win ratio and its components over time. Statistics in Medicine 38:1, 53--61.

\bibitem[{Huque {\it et al.}(2011)}]{huque2011addressing}
Huque MF, Alosh M, Bhore R. (2011). Addressing multiplicity issues of a composite endpoint and its components in clinical trials. Journal of Biopharmaceutical Statistics 21:4, 610--634.

\bibitem[{Imai and Ratkovic(2014)}]{imai2014covariate}
Imai K, Ratkovic M. (2014). Covariate balancing propensity score. Journal of the Royal Statistical Society: Series B (Statistical Methodology) 76:1, 243--263.

\bibitem[{Kim(2010)}]{kim2010calibration}
Kim JK. (2010). Calibration estimation using exponential tilting in sample surveys. Survey Methodology 36:2, 145-155.

\bibitem[{Kojadinovic and Yan(2010)}]{kojadinovic2010modeling}
Kojadinovic I, Yan J. (2010). Modeling multivariate distributions with continuous margins using the copula R package. Journal of Statistical Software 34:9, 1--20.

\bibitem[{Kurihara(2000)}]{kurihara2000traumatic}
Kurihara M. (2000). Traumatic brain injury in children. No to Hattatsu = Brain and Development 32:2, 110--115.

\bibitem[{Li {\it et al.}(2013)}]{li2013propensity}
Li Z, Alan M, Landrum MB. (2013). Propensity score weighting with multilevel data. Statistics in Medicine 32:19, 3373--3387.

\bibitem[{Lin and Ying(1994)}]{lin1994semiparametric}
Lin DY, Ying Z. (1994). Semiparametric analysis of the additive risk model. Biometrika 81:1, 61--71.

\bibitem[{Luo {\it et al.}(2017)}]{luo2017weighted}
Luo X, Qiu J, Baid S, Tian H. (2017). Weighted win loss approach for analyzing prioritized outcomes. Statistics in Medicine 36:15, 2452--2465.

\bibitem[{Luo {\it et al.}(2015)}]{luo2015alternative}
  \bibinfo{author}{Luo X, Tian H, Mohanty S, Tsai WY}. \bibinfo{year}{(2015)}.
  \bibinfo{title}{An alternative approach to confidence interval estimation for the win ratio statistic}.
  \bibinfo{journal}{Biometrics}
  \bibinfo{volume}{71}:\bibinfo{number}{1},
  \bibinfo{pages}{139--145}.

\bibitem[{Mao(2017)}]{mao2017causal}
  \bibinfo{author}{Mao L}. \bibinfo{year}{(2017)}.
  \bibinfo{title}{On causal estimation using-statistics}.
  \bibinfo{journal}{Biometrika}
  \bibinfo{volume}{105}:\bibinfo{number}{1},
  \bibinfo{pages}{215--220}.

\bibitem[{Mao(2019)}]{mao2019alternative}
 Mao L (2019). On the alternative hypotheses for the win ratio. Biometrics 75:1, 347--351.

\bibitem[{Marshall and Olkin(1988)}]{marshall1988families}
Marshall AW, Olkin I. (1988). Families of multivariate distributions. Journal of the American statistical association 83:403, 834--841.

\bibitem[{Mckinlay {\it et al.}(2008)}]{mckinlay2008prevalence}
McKinlay A, Grace RC, Horwood LJ, Fergusson DM, Ridder Elizabeth M, MacFarlane MR. (2008). Prevalence of traumatic brain injury among children, adolescents and young adults: prospective evidence from a birth cohort. Brain Injury 22:2, 175--181.

\bibitem[{Michaud {\it et al.}(1993)}]{michaud1993traumatic}
Michaud LJ, Duhaime A-C, Batshaw ML. (1993). Traumatic brain injury in children. Pediatric Clinics of North America 40:3, 553--565.

\bibitem[{Oakes(2016)}]{oakes2016win}
  \bibinfo{author}{Oakes D}.
  \bibinfo{title}{On the win-ratio statistic in clinical trials with multiple types of event}. \bibinfo{year}{(2016)}. \bibinfo{journal}{Biometrika} \bibinfo{volume}{103}:\bibinfo{number}{3}, \bibinfo{pages}{742--745},

\bibitem[{Pocock {\it et al.}(2011)}]{pocock2011win}
  \bibinfo{author}{Pocock SJ, Ariti CA, Collier TJ, Wang D}. \bibinfo{year}{(2011)}.
  \bibinfo{title}{The win ratio: a new approach to the analysis of composite endpoints in clinical trials based on clinical priorities}.
  \bibinfo{journal}{European Heart Journal}
  \bibinfo{volume}{33}:\bibinfo{number}{2},
  \bibinfo{pages}{176--182}.

\bibitem[{Rubin(1974)}]{rubin1974estimating}
Rubin DB. (1974). Estimating causal effects of treatments in randomized and nonrandomized studies. Journal of Educational Psychology 66:5, 688-701.

\bibitem[{Schennach(2007)}]{schennach2007point}
Schennach SM. (2007). Point estimation with exponentially tilted empirical likelihood. The Annals of Statistics 35:2, 634--672.

\bibitem[{Thoemmes and West(2011)}]{thoemmes2011use}
Thoemmes FJ, West SG. (2011). The use of propensity scores for nonrandomized designs with clustered data. Multivariate Behavioral Research 46:3, 514--543.

\bibitem[{Tong {\it et al.}(2012)}]{tong2012weighting}
Tong BC, Huber JC, Ascheim DD, Puskas JD, Ferguson TB, Blackstone EH, Smith PK. (2012). Weighting composite endpoints in clinical trials: essential evidence for the heart team. The Annals of Thoracic Surgery 94:6, 1908--1913.

\bibitem[{Wang and Pocock(2016)}]{wang2016win}
 Wang D, Pocock S. (2016). A win ratio approach to comparing continuous non-normal outcomes in clinical trials. Pharmaceutical Statistics 15:3, 238--245.

\bibitem[{Weintraub(2016)}]{weintraub2016statistical}
  Weintraub WS. (2016). Statistical Approaches to Composite Endpoints. JACC: Cardiovascular Interventions.

\bibitem[{Yang(2018)}]{yang2018propensity}
Yang S. (2018). Propensity score weighting for causal inference with clustered data. Journal of Causal Inference 6:2, 1-19.

\end{thebibliography}

{}

\clearpage

%\section*{Author Biography}
%
%\begin{biography}{\includegraphics[width=66pt,height=86pt,draft]{empty}}{\textbf{Author Name.} This is sample author biography text this is sample author biography text this is sample author biography text this is sample author biography text this is sample author biography text this is sample author biography text this is sample author biography text this is sample author biography text this is sample author biography text this is sample author biography text this is sample author biography text this is sample author biography text this is sample author biography text this is sample author biography text this is sample author biography text this is sample author biography text this is sample author biography text this is sample author biography text this is sample author biography text this is sample author biography text this is sample author biography text.}
%\end{biography}

\end{document}